\documentclass[a4paper]{jpconf}
\usepackage{graphicx}
\usepackage{amsmath}
\usepackage{color}
\newcommand{\refr}{}
\newcommand{\refb}{}
\newcommand{\refbb}{}
\newcommand{\refbbb}{}
\begin{document}
\title{{\refbb Single-Source} Gravitational Wave Limits From the J1713+0747 24-hr Global Campaign}

\author{ 
T.\,Dolch$^{1,2}$ (for the NANOGrav Collaboration), 
J.~A.~Ellis$^{3}$, 
S.\,Chatterjee$^{2}$,
J.\,M.\,Cordes$^{2}$,
M.\,T.\,Lam$^{2}$,
C.\,Bassa$^{4,5}$,
B.\,Bhattacharyya$^{5}$,
D.\,J.\,Champion$^{6}$,
I.\,Cognard$^{7}$,
K.\,Crowter$^{8}$,
P.\,B.\,Demorest$^{9}$,
J.\,W.\,T.\,Hessels$^{4,10}$,
G.\,Janssen$^{4}$,
F.\,A.\,Jenet$^{11}$,
G.\,Jones$^{12}$,
C.\,Jordan$^{5}$,
R.\,Karuppusamy$^{6}$, %
M.\,Keith$^{5}$,
V.~I.\,Kondratiev$^{4,13}$,
M.\,Kramer$^{6,14}$,
P.\,Lazarus$^{6}$,
T.\,J.\,W.\,Lazio$^{3}$,
D.~R.~Lorimer$^{15}$,
D.\,R.\,Madison$^{2,16}$,
M.\,A.\,McLaughlin$^{15}$,
N.\,Palliyaguru$^{15}$,
D.\,Perrodin$^{17}$,
S.~M.\,Ransom$^{16}$,
J.\,Roy$^{5,18}$,
R.\,M.\,Shannon$^{19}$,
R.\,Smits$^{4}$,
I.~H.\,Stairs$^{8}$,
B.~W.\,Stappers$^{5}$,
D.~R.~Stinebring$^{20}$, 
K.\,Stovall$^{21}$,
J.\,P.\,W.\,Verbiest$^{6,22}$,
W.\,W.\,Zhu$^{6}$}

\address{$^{1}$Department of Physics, Hillsdale College, 33 E. College Street, Hillsdale, MI 49242, USA}
\address{$^{2}$Astronomy Department, Cornell University, Ithaca, NY 14853, USA}
\address{$^{3}$Jet Propulsion Laboratory, California Institute of Technology, 4800 Oak Grove Drive, Pasadena, CA 91106, USA}
\address{$^{4}$ASTRON, the Netherlands Institute for Radio Astronomy, Postbus 2, 7990 AA, Dwingeloo, The Netherlands}
\address{$^{5}$Jodrell Bank Centre for Astrophysics, School of Physics and Astronomy, The University of Manchester, Manchester M13 9PL, UK}
\address{$^{6}$Max-Planck-Institut f\"ur Radioastronomie, Auf dem H\"ugel 69, D-53121 Bonn, Germany}
\address{$^{7}$Laboratoire de Physique et Chimie de l'Environnement et de l'Espace, LPC2E UMR 6115 CNRS, F-45071 Orl\'eans Cedex 02, and Station de radioastronomie de Nan\c{c}ay, Observatoire de Paris, CNRS/INSU, F-18330 Nan\c{c}ay, France}
\address{$^{8}$Department of Physics and Astronomy, University of British Columbia, 6224 Agricultural Road, Vancouver, BC V6T 1Z1, Canada}
\address{$^{9}$National Radio Astronomy Observatory, 1003 Lopezville Rd., Socorro, NM 87801, USA}
\address{$^{10}$Anton Pannekoek Institute for Astronomy, University of Amsterdam, Science Park 904, 1098 XH Amsterdam, The Netherlands}
\address{$^{11}$Center for Advanced Radio Astronomy, University of Texas, Rio Grande Valley, Brownsville, TX 78520, USA}
\address{$^{12}$Columbia Astrophysics Laboratory, Columbia University, NY 10027, USA}
\address{$^{13}$Astro Space Center of the Lebedev Physical Institute, Profsoyuznaya str. 84/32, Moscow 117997, Russia}
\address{$^{14}$University of Manchester, Jodrell Bank Observatory, Macclesfield, Cheshire, SK11 9DL, UK}
\address{$^{15}$Department of Physics and Astronomy, West Virginia Univ., Morgantown, WV 26506, USA}
\address{$^{16}$National Radio Astronomy Observatory, 520 Edgemont Road, Charlottesville, VA 22901, USA}
\address{$^{17}$INAF-Osservatorio Astronomico di Cagliari, Via della Scienza 5, 09047 Selargius (CA), Italy}
\address{$^{18}$National Centre for Radio Astrophysics, Tata Institute of Fundamental Research, Pune 411007, India.}
\address{$^{19}$CSIRO Astronomy \& Space Science, Australia Telescope National Facility, PO Box 76, Epping, NSW 1710, Australia}
\address{$^{20}${Department of Physics and Astronomy, Oberlin College, Oberlin, OH 44074, USA}}
\address{$^{21}$Physics and Astronomy Department, University of New Mexico, 1919 Lomas Boulevard NE, Albuquerque, NM 87131-0001, USA}
\address{$^{22}$Fakult\"at f\"ur Physik, Universit\"at Bielefeld, Postfach 100131, D-33501 Bielefeld, Germany}

\ead{tdolch@hillsdale.edu}

\begin{abstract}
Dense, continuous pulsar timing observations over a 24-hr period provide a method for probing intermediate gravitational wave (GW) frequencies from 10 microhertz to 20 millihertz. The European Pulsar Timing Array (EPTA), the North American Nanohertz Observatory for Gravitational Waves (NANOGrav), the Parkes Pulsar Timing Array (PPTA), and the combined International Pulsar Timing Array (IPTA) all use millisecond pulsar observations to detect or constrain GWs typically at nanohertz frequencies. In the case of the IPTA's nine-telescope 24-Hour Global Campaign on millisecond pulsar J1713+0747, GW limits in the intermediate frequency regime can be produced. The negligible change in dispersion measure during the observation minimizes red noise in the timing residuals, constraining any contributions from GWs due to individual sources. {\refbb At 10$^{-5}$\,Hz, the 95\% upper limit on strain is 10$^{-11}$ for GW sources in the pulsar's direction.}
\end{abstract}

\vspace*{-2em}
\section{Introduction}
The direct detection of gravitational waves (GWs) is a goal for which pulsar timing arrays (PTAs) are well-suited. PTAs are collections of millisecond pulsars (MSPs) distributed throughout the sky that are regularly monitored with large, typically single-dish radio telescopes. {\refr MSPs are recycled neutron stars (NSs) with low spin-down rates.} The rotational regularity of MSPs means that their beamed radio emission functions as a stable clock. Deviations from regularity may signify a GW-induced distortion in the spacetime metric, changing the apparent distance between pulses and thus changing the observed spin frequency of a pulsar. PTA-oriented scientific collaborations around the world (the EPTA collaboration \cite{k13}, NANOGrav \cite{mc13}, and the PPTA collaboration \cite{h13}; working together as the IPTA collaboration \cite{ma13}) have been regularly monitoring their respective {\refbb sets} of pulsars for over a decade. The best-case GW detection would show {\refr the correlated signature between the pulse times-of-arrival (TOAs) from different pulsars predicted in \cite{hd83}}. Such a signal would be free of spurious effects along a particular line-of-sight (LoS), {\refr such as long-term, stochastic rotational fluctuations in a particular pulsar, also known as timing noise or spin noise}. Although GW detection has not yet occurred with PTAs, constraints on galaxy evolution models have already been obtained \cite{s13,l15,a15c}, in some cases limiting the presence of GWs {\refr using multiple-pulsar data, and in others using data from a} single pulsar. These {\refb non-detection} limits are at frequencies near a nanohertz.

\section{Data Reduction, Pipeline, and Limits on Individual Sources}

The PSR~J1713+0747 24-hour global campaign \cite{d14} was a day-long, continuous observation using nine radio telescopes on 22-Jun 2013 (MJD 56465 -- 56466). PSR~J1713+0747 was chosen because of the low {\refbbb root-mean-squared (rms) timing residuals} with respect to its timing model, and because it is regularly observed by all the PTA collaborations. The Arecibo, Effelsberg, GMRT, Green Bank, LOFAR, Lovell, Nan\c cay, Parkes, and WSRT radio telescopes were all used for the observation.
For the much longer datasets usually used in PTAs, red noise in dispersion measure (DM) variations propagates into timing residuals \cite{c13}. DM is the quantity that measures the integrated column density of free electrons along the LoS to a pulsar. The 24-hr campaign obtained high {\refbbb signal-to-noise (S/N)} TOAs at 1.4\,GHz, showing that the timing residuals were free of any red noise due to DM variations. Noise in the residuals due to interstellar scintillation (ISS) was also modeled and shown to be white. This procedure was performed in \cite{d14} using the {\refbbb frequency-dependent (FD)} parameters \cite{a15b} that modeled the pulse profile's evolution with frequency. Otherwise, performing a weighted sum will generate non-white noise structures due the the fact that ISS creates a varying S/N across the observing band. {\refr In that case, the varying S/N would result in different TOAs in each sub-band due to the intrinsic profile's changing shape with frequency.}

The timing residuals across the varying observation bandwidths are shown as a weighted sum, {\refr which we refer to here as the ``broadband''} residuals, in Fig. 1. {\refr Around 8 hr, ISS was occurring, which decreased the S/N. The lower S/N ratio was in spite of the fact that the GMRT's sensitivity is comparable to that of other telescopes.} These {\refbb otherwise high-S/N} data motivate a search for high-frequency GWs from individual (or continuous wave; CW) merging supermassive black hole binaries (SMBHBs), rather than a stochastic background of GWs from such sources from 10$^{-9}$ -- 10$^{-7}$\,Hz. For the 24-hr campaign, we are sensitive to GWs at 10$^{-5}$~--~10$^{-3}$\,Hz. At these high GW frequencies, we do not expect sensitivity to the GW background because of its powerlaw spectrum, which dominates at low frequencies. However, a nearby SMBHB with a sufficient chirp-mass could produce CWs. \cite{c14} also demonstrates that the 10$^{-5}$ -- 10$^{-3}$\,Hz range may be a good discovery band for unanticipated, periodic GW sources. {\refb We use} the same detection procedure for sinusoidal GWs as \cite{a14}, where details of the pipeline can be found. The only difference between the pipeline as typically used for a PTA dataset and for the present dataset is simply that only one pulsar, PSR~J1713+0747, is included in the array. \cite{a14} contains several detection algorithms; here we utilize the frequentist, fixed-noise model, given that we have already established the white-noise character of the dataset. The pipeline accounts for the uneven sampling and the varying S/N of the data over time. {\refr An additional pulse phase term due to any GW perturbations on the pulsar itself is included.}
\vspace*{-1.5em}

\begin{figure}[h]
\begin{minipage}{18pc}
\hspace*{-1em}
\includegraphics[width=20pc]{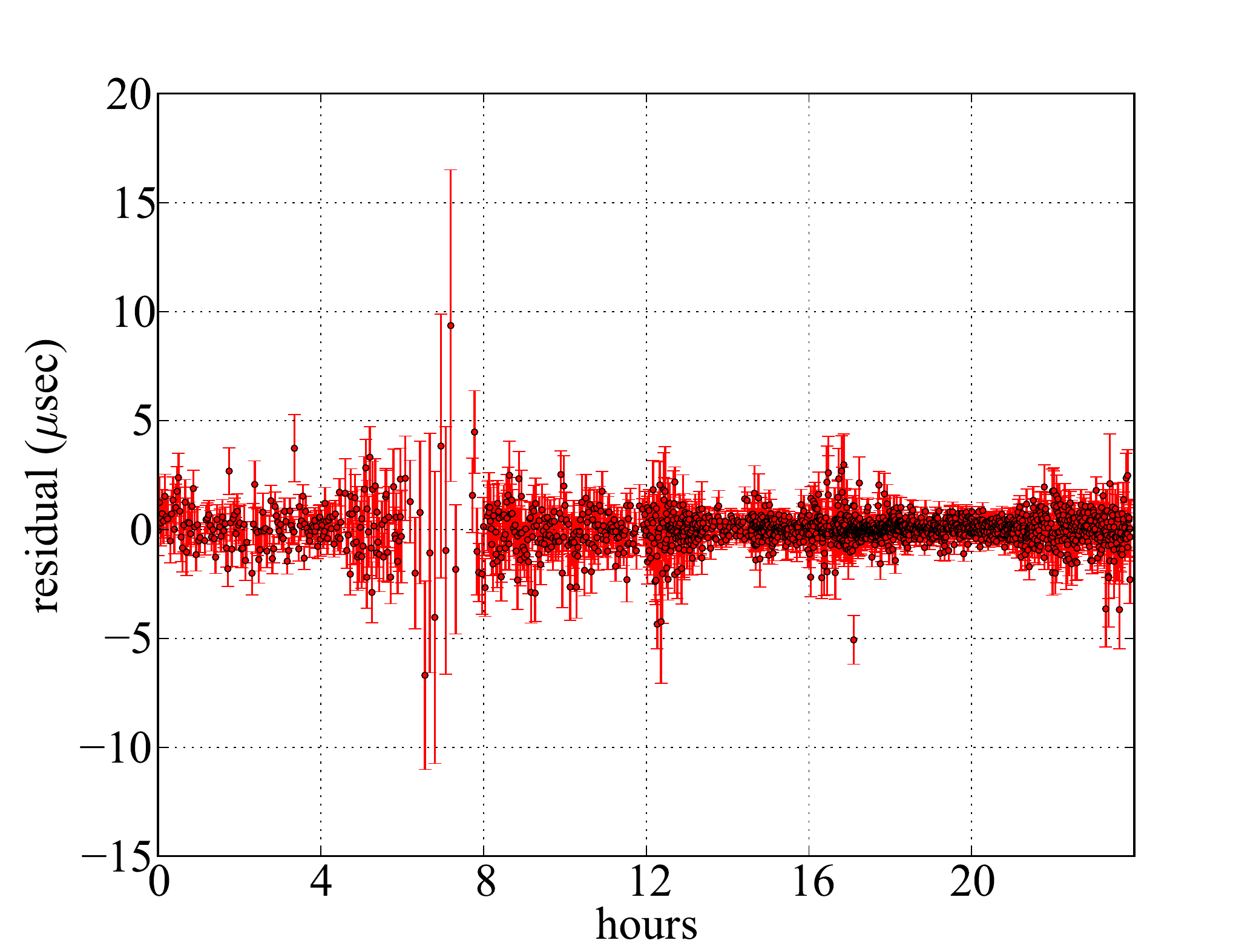}
\vspace*{-2em}
\caption{\label{label}The 1.4\,GHz broadband, 8-telescope timing residuals for the 24-hour global campaign on pulsar J1713+0747. The white-noise character of these residuals motives a search for GW in the frequency range 10$^{-5}$ -- 10$^{-3}$\,Hz. {\refr The integration time is 120\,s, the variation in rms is due to ISS and changing observation bandwidth, and the error bars are the 1\,$\sigma$ residual uncertainty.}}
\end{minipage}
\hspace{1em}
\begin{minipage}{18pc}
\hspace{-2.5em} 
\vspace{4.5em} 
\includegraphics[width=22pc]{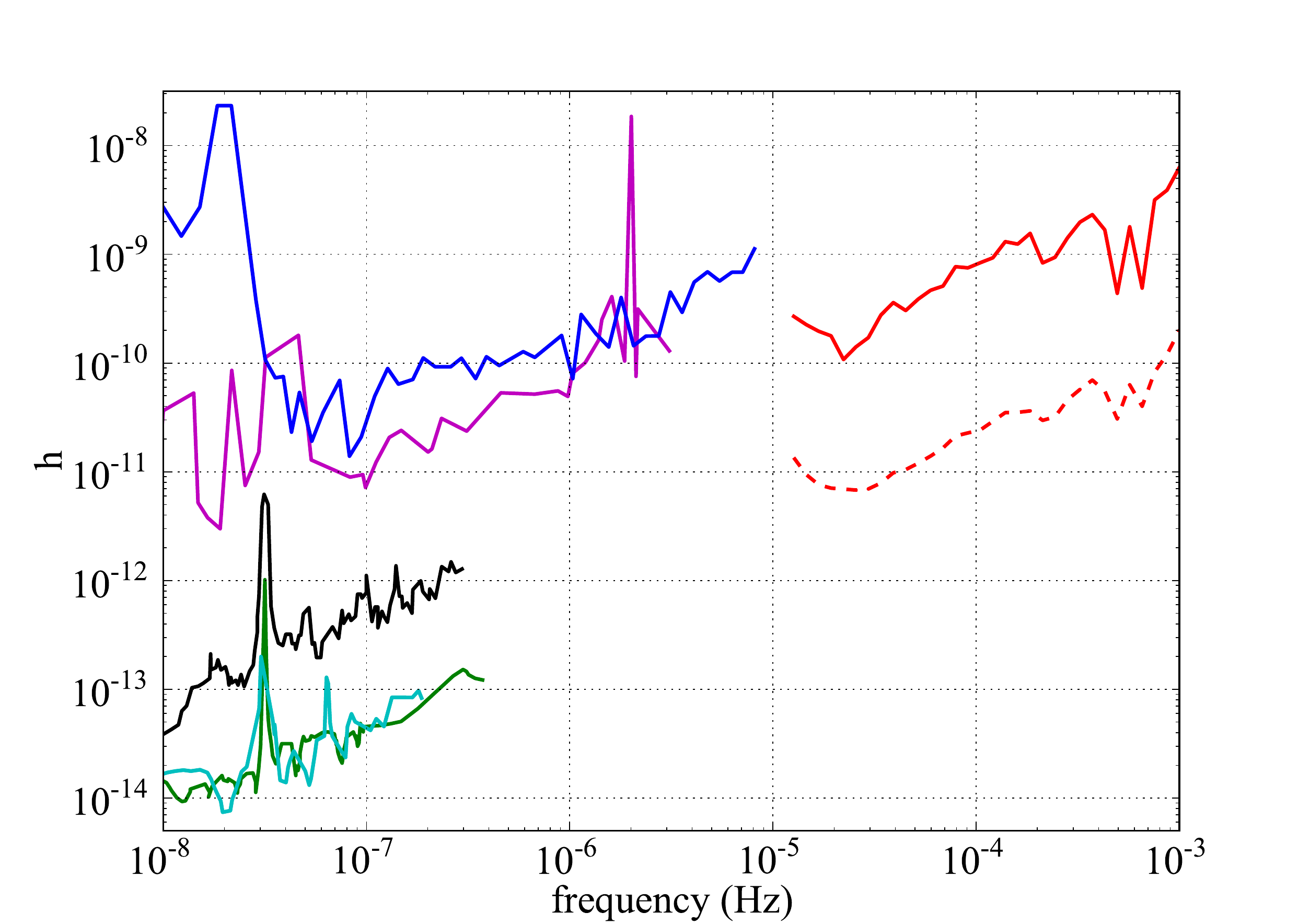}
\vspace*{-5.5em}
\caption{\label{label}GW limits from the J1713 24-hr global campaign (red) from individual GW sources in random sky directions. The dashed curve shows the GW limit in the direction of the pulsar. The other CW limits shown are: pulsar J0437 \cite{y10} (blue), pulsar B1937 \cite{y14} (magenta), NANOGrav \cite{a14} (black), the EPTA \cite{b15} (green), and the PPTA \cite{z14} (cyan).}
\end{minipage} 
\end{figure}

We obtain two {\refb 95\% upper limit non-detection} curves (Fig. 2) for the 24-hour global data: one for individual sources in a random direction in the sky (solid red curve), and the other for sources in the direction of PSR~J1713+0747 itself (dashed red curve). Both show limits in GW strain $h$ as a function of GW frequency ($f_{\textrm{GW}}$). We chose a 0.01 radian cone for the directional limit. A spin-down rate fit is also included. The directional limit curve, {\refbb which has $h\sim10^{-11}$ at 10$^{-5}$\,Hz,}  provides about a factor of 10 improvement. Also plotted are the single-pulsar high-frequency limit curves from \cite{y10} and \cite{y14}, from PSR~J0437--4715 (magenta) and PSR~B1937+21 (blue), respectively, {\refr for CW sources from a random sky direction}. The random-direction CW limit curve from \cite{a14} (black) is also shown, even though it is derived from 17 NANOGrav pulsars rather than from a single pulsar. The curve is included because it was generated from the same CW detection pipeline; many-pulsar CW limits from the EPTA \cite{b15} (green) and the PPTA \cite{z14} (cyan), both random-direction, are also shown. We note that the NANOGrav curve can be extrapolated to meet the upper curve from the 24-hr campaign. This is consistent with the fact that PSR J1713+0747 influences the GW search procedure across all pulsars because of its exceptionally high residual S/N. {\refr Both \cite{hui13} and the Cassini radio science investigation \cite{a03} also derived limits on GWs in similar $f_{\textrm{GW}}$ bands, but these apply only to a stochastic background and we do not consider them here. There are no known GW sources near this band that are strong enough to have been detected. According to \cite{nsref}, the double pulsar J0737--3039 1.1\,kpc away \cite{b03,l04,d09} should produce $h_{\textrm{max}}\sim$\,10$^{-21}$ at $f_{\textrm{GW}}\sim$\,0.2\,mHz. {\refbbb Indeed, the current GW frequency of J0737--3039 is the highest known amongst NS binaries, and is closest to the GW frequency band considered in this work.} Further considerations about other potential GW sources will be explored in a future publication. Meanwhile, we recommend further long-track studies, especially those optimized for different sky directions, or those coordinated between different telescopes monitoring different pulsars simultaneously. 

\section{Acknowledgments}

The work of SC, JMC, PBD, TD, JAE, FJ, GJ, MTL, TJWL, DRL, DRM, MAM, NP, SMR, DRS and KS was partially supported through the NSF-PIRE program award number 0968296 and NSF-PFC award number 1430284. TD acknowledges seed-funding from the NANOGrav PFC award. PL acknowledges the support of IMPRS Bonn/Cologne and FQRNT B2. Some of this work was supported by the ERC Advanced Grant ``LEAP'', Grant Agreement Number 227947 (PI M. Kramer). Part of this research was carried out at the Jet Propulsion Laboratory, California Institute of Technology, under a contract with NASA. JAE acknowledges support by NASA through Einstein Fellowship grant PF4-150120. LOFAR data analysis was supported by the European Research Council under the EU's Seventh Framework Programme (FP/2007-2013) / ERC Grant Agreement \#337062 (PI J. Hessels). {\refbbb We thank the anonymous referee.}

\smallskip

\end{document}